\documentclass[3p,times,twocolumn]{elsarticle}
\newcommand{\be}{\begin{equation}}
\newcommand{\ee}{\end{equation}}

\usepackage{ecrc}

\volume{00}

\firstpage{1}

\journalname{Nuclear Physics B Proceedings Supplement}

\runauth{F. Scardina et al.}

\jid{nppp}

\jnltitlelogo{Nuclear Physics B Proceedings Supplement}

\usepackage{amssymb}

\usepackage[figuresright]{rotating}

\begin{document}

\begin{frontmatter}




\title{Toward an understanding of the $R_{AA}$ and $v_2$ puzzle for heavy quarks}


\author{F. Scardina $^{1,2}$, S. K. Das$^{1,2}$, S. Plumari$^{1,2}$, J.,I. Bellone$^{1,2}$, 
and V. Greco$^{1,2}$}

\address{$^1$ Department of Physics and Astronomy, University of Catania, Via S. Sofia 64, I-
95125 Catania, Italy}
\address{$^2$ Laboratori Nazionali del Sud, INFN-LNS, Via S. Sofia 62, I-95123 Catania, Italy}

\ead{scardinaf@lns.infn.it}

\begin{abstract}
One of the primary aims of the ongoing nuclear collisions at Relativistic Heavy Ion Collider (RHIC) 
and Large Hadron  Collider (LHC) energies is to create a Quark Gluon Plasma (QGP). 
The heavy quarks constitutes a unique probe of the QGP properties. 
Both at RHIC and LHC energies a puzzling relation between the nuclear modification factor $R_{AA}(p_T)$ 
and the elliptic flow $v_2(p_T)$ related to heavy quark has been observed which challenged all
the existing models.\\
We discuss how the temperature dependence of the heavy quark drag coefficient can
address for a large part of such a puzzle. We have considered four different 
models to evaluate the temperature dependence of drag and diffusion coefficients propagating through
a quark gluon plasma (QGP). All the four different models are set to reproduce the same $R_{AA}(p_T)$ 
experimentally observed at RHIC energy. We have found  that for the same $R_{AA}(p_T)$ one can 
generate $2-3$ times more $v_{2}$ depending on the temperature dependence of 
the heavy quark drag coefficient.
\end{abstract}




\end{frontmatter}


\section{Introduction}
\label{Intro}
Heavy Quarks, charm and bottom, created in ultra-Relativistic Heavy Ion Collisions (uRHIC) 
represents ideal probes to study  the Quark Gluon Plasma QGP \cite{BS,hfr}.
An essential feature in analyzing Heavy quarks motion in a QGP
is that their mass is much larger than  the typical momentum exchanged with the
plasma particles entailing that many soft scatterings are necessary to change
significantly the momentum and the trajectory of the heavy quarks.
Therefore the propagation of heavy quarks has been usually treated as a Brownian
motion that is described by means of the Fokker-Planck (FP) equation. In such an equation the interaction 
is encoded in the drag and diffusion coefficient. 
The two key observables related to HQ that have been measured in experiments are 
the nuclear suppression factor  $R_{AA}$ and the elliptic flow $v_2$ 
\cite{phenixelat,stare, adare2,alice,alice2}. 
Several theoretical efforts have been made using the Fokker Planck equation to  reproduce 
the $R_{AA}$ and the  $v_2$ experimentally observed 
\cite{DKS,rappv2,Das,alberico,bass, hees,ali,Das:2015ana}
Another approach used to describe heavy quark propagation is the 
Boltzmann approach (BM)\cite{gossiauxv2,gre,fs,fs2,you,Song:2015sfa}. 
Analyzing the $R_{AA}$ gives information on the average magnitude of the interaction. 
We point out that studying the relation between $R_{AA}$ and $v_{2}$ it is possible to deduce
other informations on the interactions between HQ and bulk.
All the approaches show some difficulties to describe simultaneously $R_{AA}$  and $v_2$.
We have found that two ingredients assume a particular importance in reducing the differences 
between experimental data for  $R_{AA}$  and $v_2$ and theoretical calculations: the temperature 
dependence of the interaction and the use of full Boltzmann collision integral 
to study the time evolution of HQ momentum.
The proceeding is organized as follows. In section 2 we discuss the Fokker Planck 
approach which is used to describe the propagation of heavy quark
through the QGP. In this section we present four different modelings to calculate the drag and 
diffusion coefficients.
In section 3  we compare the result for $R_{AA}$ and  $v_2$ obtained with FP approach with those obtained using 
the BM approach.
\section{Fokker-Planck approach}
\begin{figure}[ht]
\includegraphics[width=16pc]{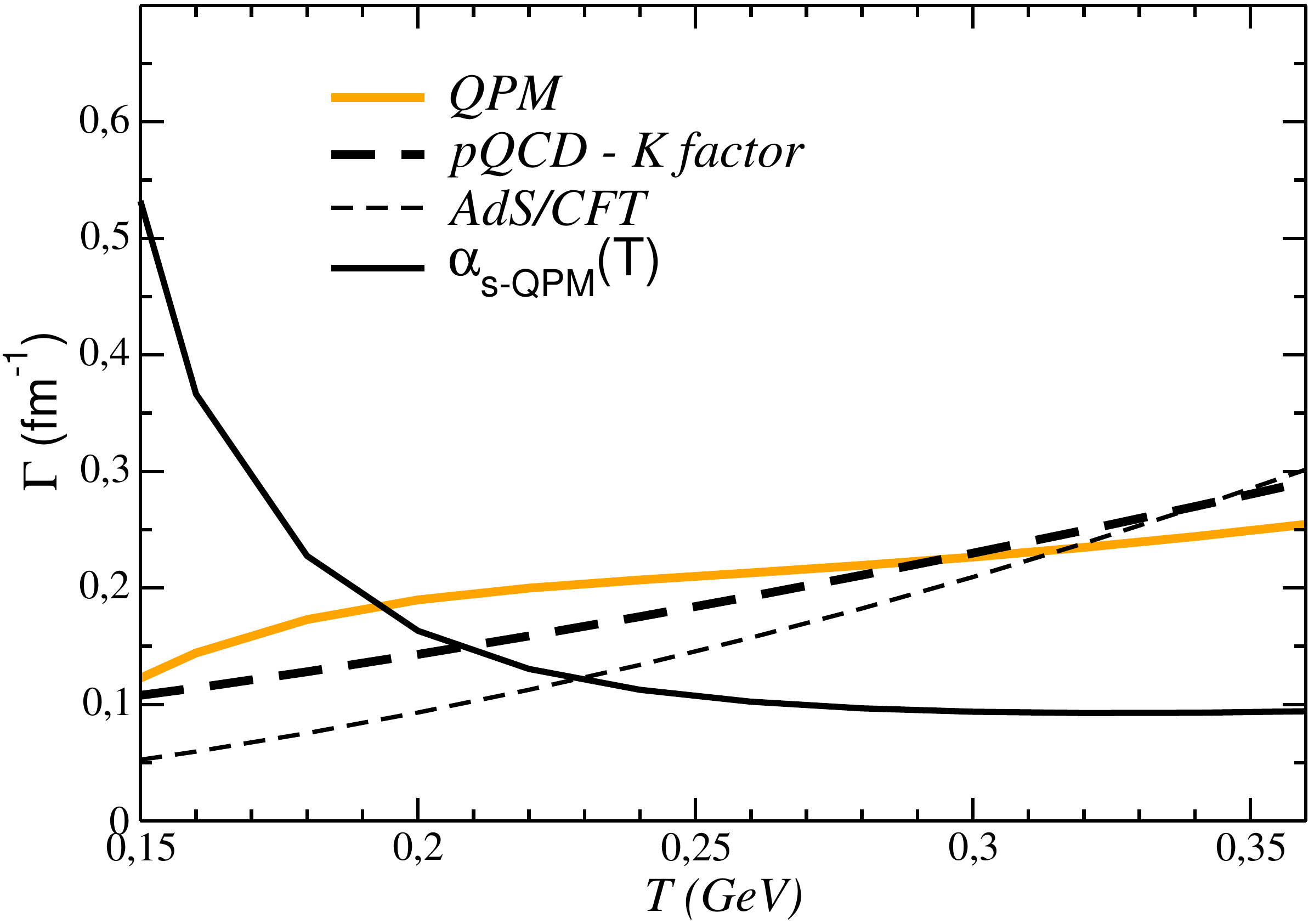}
\caption{Variation of the drag coefficients with respect to temperature at $p=0.1 GeV$}
\label{drag}
\end{figure}
The propagation of the Heavy quark is described by the Fokker Planck equations  which is indicated
in the following equation 
\begin{eqnarray}
\frac{\partial f}{\partial t}=\frac{\partial}{\partial p_i} \left[ A_i({\bf p}) f 
+ \frac{\partial}{\partial p_j}[B_{ij}(\bf p)]\right]
\label{eq:FP}
\end{eqnarray}
where  $ A_i$ and $B_{ij}$ are the  drag and  the diffusion coefficients.
To study the propagation of HQ this equation is replaced by  the 
relativistic Langevin equation, which is more suited for numerical simulations
\begin{eqnarray}
dx_i &=&\frac{p_i}{E}dt, \nonumber \\
dp_i &=&-A p_i dt+(\sqrt{2B_0}P_{ij}^{\perp}+\sqrt{2B_1}P_{ij}^{\parallel}) \rho_j\sqrt{dt}
\label{L_V}
\end{eqnarray}
\begin{figure}[ht]
\includegraphics[width=16pc]{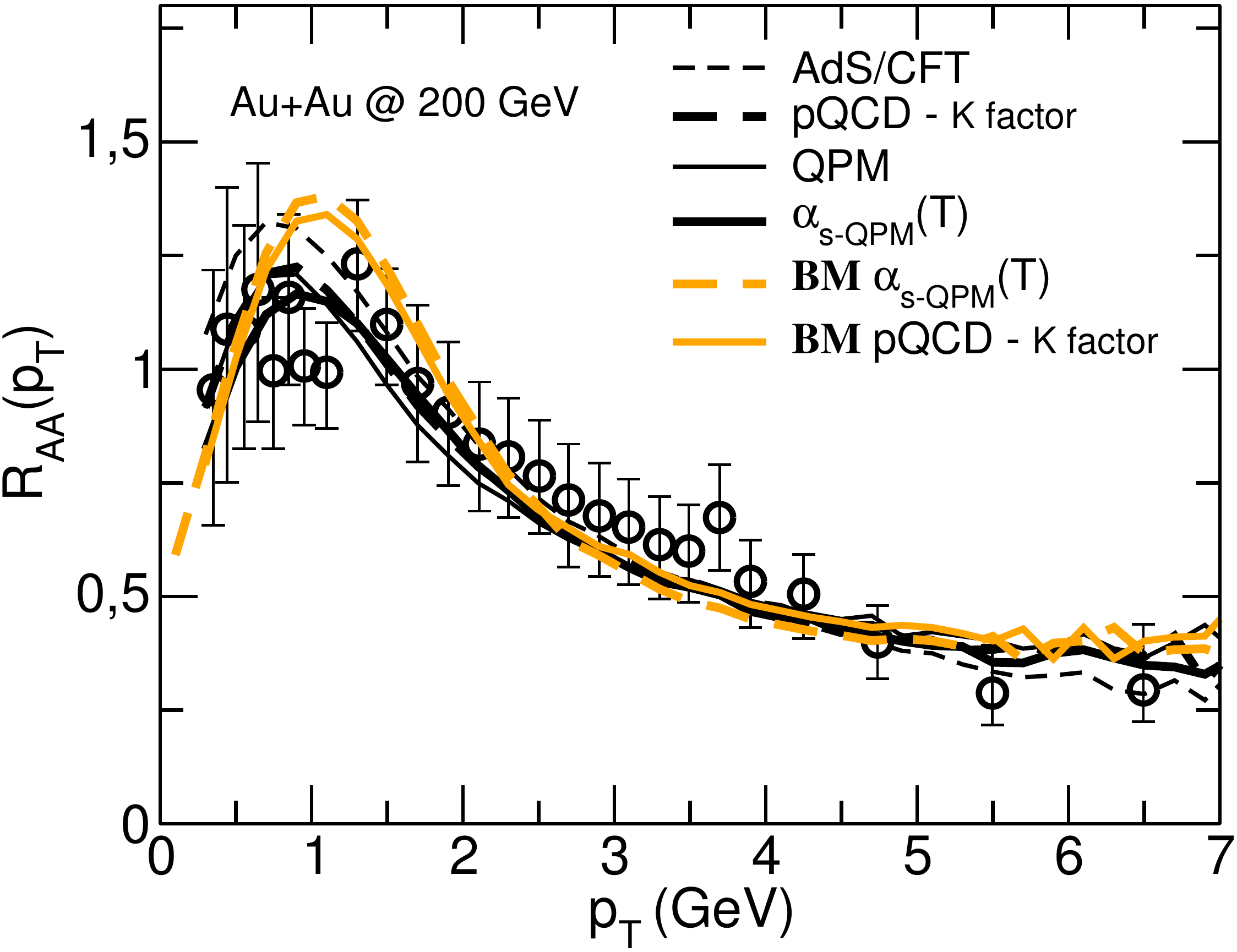}
\caption{Comparison of the experimental data for the  nuclear suppression factor  as a function
of $p_T$ at RHIC with the results we get  with the FP equation for the four different
T-dependences of the drag coefficient (Black lines). The orange lines show the results we get using the 
Boltzmann approach for the pQCD case and  for the $\alpha_{QPM}(T),m_q=m_g=0$ case.}
\label{RAA_RHIC}
\end{figure}
\begin{figure}[ht]
\includegraphics[width=16pc]{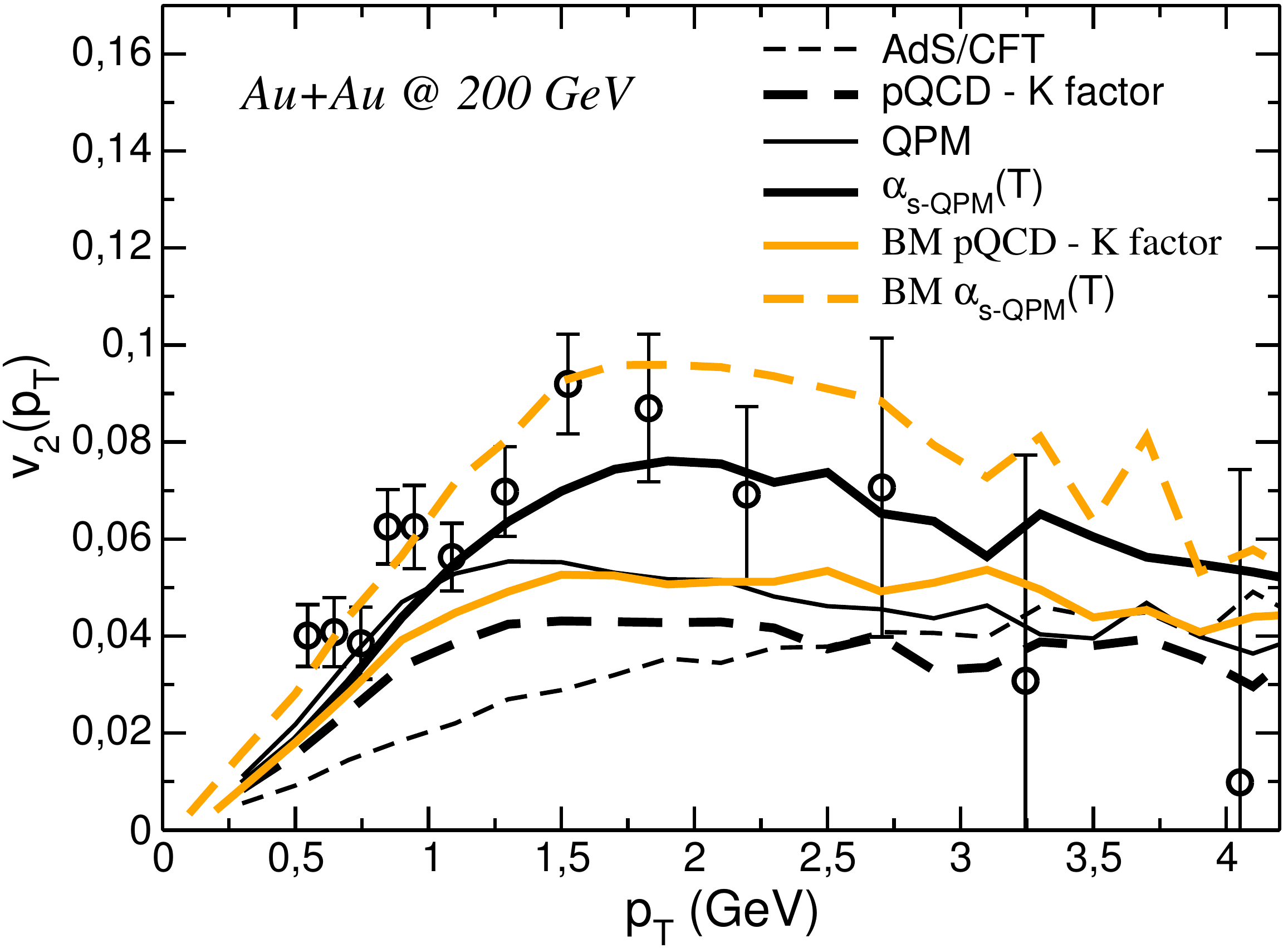}
\caption{Comparison of the experimental data for the  elliptic flow  as a function
of $p_T$ at RHIC with the result we get  with the FP equation for the four different
T-dependences of the drag coefficient (Black lines). The orange lines show the results we get using the 
Boltzmann approach for the pQCD case and  for the $\alpha_{QPM}(T),m_q=m_g=0$ case.}
\label{v2_RHIC}
\end{figure}
where $dx_i$ and $dp_i$ are the coordinates and momenta changes in each time step $dt$;
$A$ is the drag force and $B_0$ and $B_1$ are respectively the longitudinal and the transverse 
diffusion coefficients; $\rho_j$
is the stochastic Gaussian distributed variable; $P_{ij}^{\perp}= \delta_{ij}-p_i p_j/p^2$ and 
$P_{ij}^{\parallel}=p_i p_j/p^2$ 
are the transverse and longitudinal projector operators respectively.
We employ the common assumption, $B_0=B_1=D$ \cite{rappv2,Das,bass,hees}.
To solve the Langevin equation a background medium which describes the evolution 
of the bulk is necessary. 
In our work the  
evolution of the bulk is provided by a 3+1D relativistic transport code tuned 
at fixed $\eta/s$ ~\cite{greco_cascade,plumari:2012prc} which is able 
to reproduce the same results of hydrodynamical  simulations.
The transport code provides at each time step the density profile and the temperature profile 
of the bulk which  are necessary to calculate the drag coefficients.
We have performed simulations of $Au+Au$ collisions 
at $\sqrt{s}= 200$ AGeV for the minimum bias. 
In both cases  the initial conditions in coordinate space are given by the  Glauber 
model, while in the momentum space a Boltzmann-Juttner distribution function 
up to a transverse momentum $p_T=2$ GeV has been considered. 
At larger momenta mini-jet distributions, as calculated within pQCD at NLO order, 
have been employed \cite{Greco:2003xt}. 
The HQ distribution in momentum space is in accordance with the one in proton-proton that 
have been taken from \cite {Cacciari:2005rk}.
We have used four different modelings to calculate the drag coefficient 
entailing  different temperature dependence of the interaction . 
The diffusion coefficient is instead calculated  in accordance  with the Einstein  
relation $D=\Gamma E T$. Our purpose is to investigate the relation between $R_{AA}$  and  $v_{2}$ 
for  different temperature dependence of the energy loss.
In the first modeling we have evaluated the drag coefficient from (pQCD)
and we have considered elastic interaction among HQ and the bulk (light quarks and gluons).
The scattering matrix related to these processes  
${\cal M}_{gHQ}$, ${\cal M}_{qHQ}$ and ${\cal M}_{\bar{q}HQ}$  
in leading order are the well known Combridge matrix.
The  infrared singularity in the $t$-channel
is regularized introducing a Debye screening mass 
$m_D=\sqrt{4\pi\alpha_s}\, T$ with a running coupling \cite{zantow}.\\
In another modeling we have evaluated the drag force from the gauge/string
duality~ \cite{Maldacena:1997re} through the following equation
\be
\Gamma_{conf}= C \frac{T_{QCD}^2}{M_c} \\
\label{confdrag}
\ee
where $C={\pi\sqrt\lambda\over 2\sqrt3}=2.1\pm0.5$\cite{ali}.\\
We have also considered another modeling in which 
the drag coefficient is evaluated  considering a bulk consisting of particles with a 
T-dependent quasi-particle masses, $m_q=1/3g^2T^2$, $m_g=3/4g^2T^2$.
This model  is able to reproduce the thermodynamics 
of lattice QCD~\cite{salvo} (see also ~\cite{vc,elina,elina2})
by fitting the coupling $g(T)$. Such a fit leads to the following coupling~\cite{salvo}: 
\be
g^2(T)=\frac{48\pi^2}{[(11N_c-2N_f)ln[\lambda(\frac{T}{T_c}-\frac{T_s}{T_c})]^2}
\label{running_g}
\ee
where $\lambda$=2.6 and $T/T_s$=0.57.\\ 
Finally we have considered a model  
in which the light quarks and gluons are massless but the coupling is from 
the QPM as indicated in Eq. \ref{running_g}. This last case is indicated in the figure as 
($\alpha_{QPM}(T),m_q=m_g=0$) and has to be 
considered as an expedient to have a drag which decreases  with the temperature.\\ 
For all the four cases considered the interaction has been rescaled  to reproduce the
$R_{AA}$ observed in experiments.
We have simulated HQ propagation with  
the Langevin dynamics for the four different models presented above. The Langevin 
equation gives as output the momentum distributions of HQ at the quark-Hadron transition 
temperature $T_c$. The momenta distributions are convoluted with the  Peterson fragmentation 
functions  of the heavy quark indicated in Eq. \ref{Peterson} in order to get the momentum 
distribution of D and B mesons.
\be
f(z) \propto 
\frac{1}{\lbrack z \lbrack 1- \frac{1}{z}- \frac{\epsilon_c}{1-z} \rbrack^2 \rbrack}
\label{Peterson}
\ee
where $\epsilon_c=0.04$ for charm quarks and  $\epsilon_c=0.005$ for bottom quark.
In figure \ref{RAA_RHIC} the nuclear modification factor $R_{AA}$ of the D and B mesons   
is shown as a function of $p_T$ for RHIC ($200 AGeV$). Instead in 
figure  \ref{v2_RHIC} the elliptic flow ($v_2=\langle ((p_x^2-p_y^2)/(p_x^2+p_y^2)) \rangle$) at the 
same energy  as a function of $p_T$ is depicted.
We observe that  the larger is the interaction in the region of low temperature 
the larger is the elliptic flow. The same conclusions has been discussed  also in the 
light flavor sector as shown in Refs. \cite{Scardina:2010zz,Liao:2008dk}.
The reason of such a strong dependence of the elliptic flow on the temperature dependence of the 
drag coefficient  is due to the fact that the elliptic flow is generated in the final stage of 
the evolution of the fireball when the temperature is lower. 
\section{Boltzmann approach}
The Boltzmann equation for the HQ distribution function is indicated here 
\begin{equation}
 p^{\mu} \partial_{\mu}f_{Q}(x,p)= {\cal C}[f_{Q}](x,p)
\label{B_E}
\end{equation}
where ${\cal{C}}[f_{Q}](x,p)$ is the relativistic Boltzmann-like collision integral
which is solved by means of a stochastic algorithm. In such an algorithm whether 
a collision happen or not is sampled stochastically comparing the collision probability 
$P_{22}=v_{rel}\sigma_{g,q+Q \rightarrow g,q+Q}\cdot\Delta t/\Delta x$
with a random number extracted between $0$ and $1$~
\cite{Lang,greco_cascade,scardina:2014prc,Greiner_cascade}.
We use the Boltzmann equation to describe the propagation of the heavy quark and  the evolution 
of the bulk\\
The comparison  between LV and BM approach  has been  studied 
in these references \cite{fs,fs2} where it is shown that for charm quark FP deviates significantly 
from the BM and such a deviation significantly  
depends on the the values of the Debye screening mass, whereas for bottom quarks 
the FP is a very good approximation. We considered in references  \cite{fs,fs2} 
three values of $m_D$: 0.4 GeV, 0.83 GeV and 1.6 GeV.
Here we have not considered a fixed value of the Debye screening mass but a value which depends on 
the temperature according to $m_D=gt$. 
\begin{figure}[ht]
\includegraphics[width=16pc]{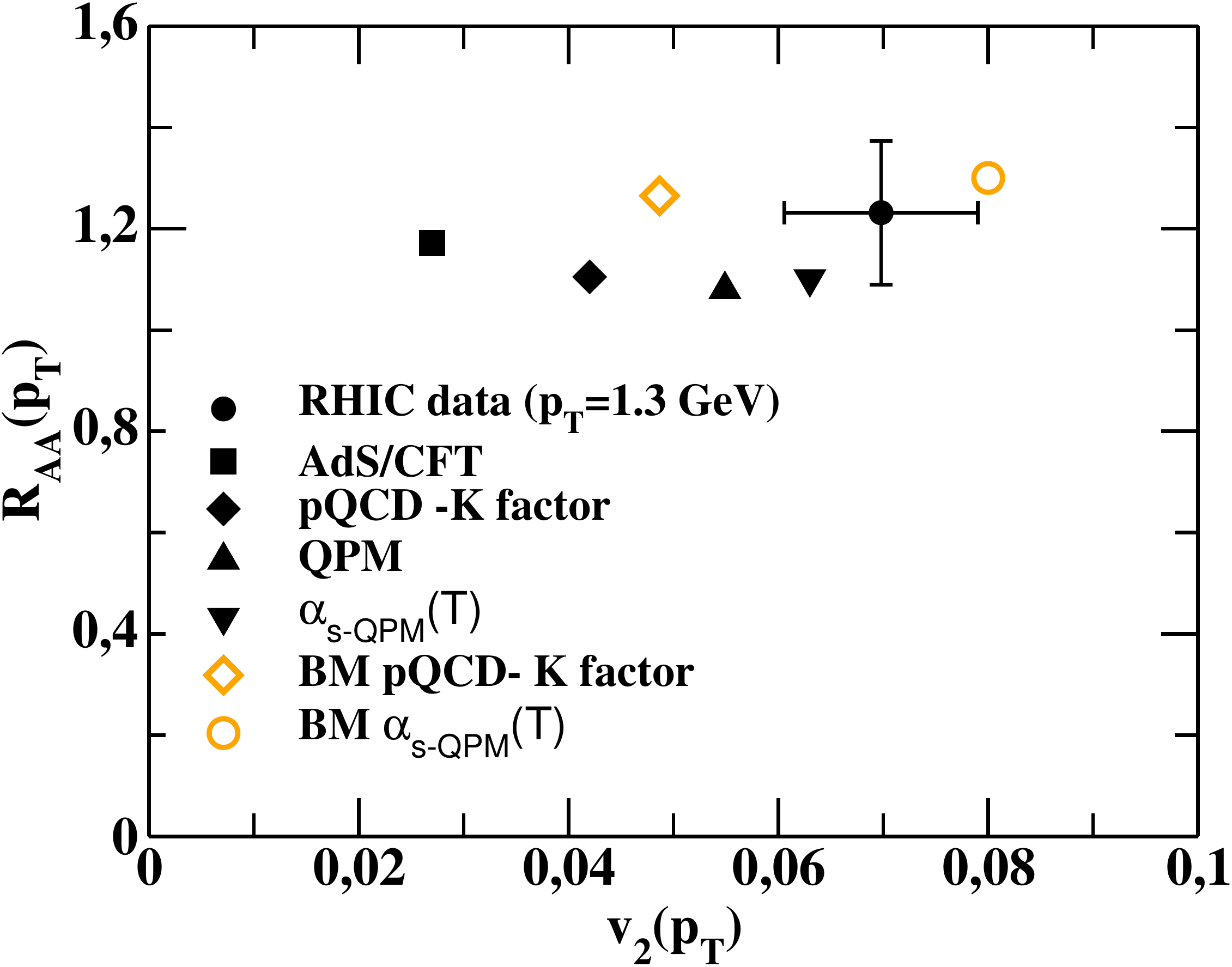}
\caption{ $R_{AA}$  vs $v_2$ obtained with the FP for the four 
different T-dependences of the drag coefficient with the experimental data at RHIC energy 
at $p_T=1.3 GeV$. The open symbols indicate the results obtained using the BM approach}
\label{RAA_vs_v2}
\end{figure}
In figures \ref{RAA_RHIC} and \ref{v2_RHIC} the comparison for the $R_{AA}$  and $v_2$ at RHIC 
between the BM (orange lines) and the FP (black lines)  are shown.
We found that using the BM for the same values of the $R_{AA}$ we get larger values for   
$v_{2}$ with respect to those obtained using the FP. \\
With the Boltzmann approach using the ($\alpha_{QPM}(T),m_q=m_g=0$) we get a value 
of the elliptic flow even larger 
with respect to the experimental data, however this case 
represent an extreme and not realistic case.  
Our results show  a non-negligible impact of the approximation
in the collision integral involved in the Fokker Plack equation on
the relation between $R_{AA}$ and $v_2$.
We summarize the results introducing a new plot  in Fig. \ref{RAA_vs_v2} in which $R_{AA}$ vs $v_2$
at a given momentum ($p_T=1.3$ GeV) is shown.  This figures clearly shows how  
the  building up of the $v_2$ can differ up to a factor 3 for the same $R_{AA}$ depending on the 
temperature dependence of the interaction and on the approach, BM or FP, used to describe the 
propagation of the heavy quark in the QGP.\\ 
\section{Conclusions}
We have found that different temperature dependences of the interaction  can lead to difference 
in the elliptic flow $v_2$ by 2-3 times even if leading  to the  same $R_{AA}$. Our studies suggest that 
the correct temperature 
dependence of the drag coefficient may not be larger power of $T$ (as in pQCD or AdS/CFT) 
rather a lower power of T or may be constant in T.\\
Moreover we have studied the difference in the building up of the elliptic flow 
between Fokker-Planck and 
Boltzmann approach for a fixed $R_{AA}$. We observe that the Boltzmann 
approach generates a larger elliptic flow  than the Fokker-Planck
\cite{fs,fs2}.

\section*{Acknowledgements} 
F. Scardina, S. K. Das, S. Plumari and V. Greco acknowledge the support by the ERC StG under the QGPDyn
Grant n. 259684

\end{document}